\documentclass[prd,aps,twocolumn,amsmath,amssymb,nofootinbib,preprintnumbers]
{revtex4}

\usepackage{graphicx}
\usepackage{dcolumn}
\usepackage{bm}
\usepackage{amsmath}
\usepackage{amsfonts}
\usepackage{euscript,bbm}
\usepackage{ifthen}
\usepackage{psfrag}
\usepackage{slashed}

\def\ls{\mathrel{\lower4pt\vbox{\lineskip=0pt\baselineskip=0pt
           \hbox{$<$}\hbox{$\sim$}}}}
\def\gs{\mathrel{\lower4pt\vbox{\lineskip=0pt\baselineskip=0pt
           \hbox{$>$}\hbox{$\sim$}}}}
\def\drawbox#1#2{\hrule height#2pt

\hbox{\vrule width#2pt height#1pt \kern#1pt
              \vrule width#2pt}
              \hrule height#2pt}

\def\Asym#1#2{\vcenter{\vbox{\drawbox{#1}{#2}
              \kern-#2pt       
              \drawbox{#1}{#2}}}}

\newcommand{\be}{\begin{equation}}
\newcommand{\ee}{\end{equation}}
\newcommand{\bea}{\begin{eqnarray}}
\newcommand{\eea}{\end{eqnarray}}

\begin{document}

\title{Natural GeV Dark Matter and the Baryon-Dark Matter Coincidence Puzzle}

\author{Rouzbeh Allahverdi$^{1}$}
\author{Bhaskar Dutta$^{2}$}

\affiliation{$^{1}$~Department of Physics and Astronomy, University of New Mexico, Albuquerque, NM 87131, USA \\
$^{2}$~Mitchell Institute of Fundamental Physics and Astronomy, Department of Physics and Astronomy, Texas A\&M University, College Station, TX 77843-4242, USA}

\begin{abstract}
We present a simple extension of the standard model that gives rise to baryogenesis a has a dark matter candidate of ${\cal O}({\rm GeV})$ mass. A minimal set of new fields required for baryogenesis includes two ${\cal O}({\rm TeV})$ colored scalars and a singlet fermion. The fermion also becomes a viable dark matter candidate when it is nearly degenerate in mass with the proton. Dark matter and baryon asymmetry are produced form the decay of heavy scalars, which can lead to a natural explanation of the baryon-dark matter coincidence problem. The dark matter candidate escapes direct and indirect detection, but can be probed at the LHC. The supersymmetric extension of this model is straightforward and leads to a multi-component dark matter scenario, which improves the direct and indirect detection prospects.
\end{abstract}
MIFPA-13-12 \\ April, 2013
\maketitle

\section{Introduction}

Weakly interacting massive particles (WIMPs) are promising dark matter (DM) candidates~\cite{WIMP}. WIMPs typically arise in models of particle physics beyond the standard model (SM). In supersymmetric (SUSY) models with conserved $R$-parity, the lightest supersymmetric particle (LSP) is the DM candidate. In the standard scenario, the DM relic abundance, precisely measured by cosmic microwave background experiments~\cite{WMAP}, is explained via thermal freeze-out of LSP annihilation in the early universe.

However, the current LHC bounds~\cite{LHC} have put this scenario under increasing pressure. The fact that no SUSY particles have been found so far, keeps pushing up the limits on the mass of gluinos and squarks of the first two generations, and to a lesser extent the stop and sbottom. This has motivated new scenarios, like natural SUSY~\cite{Natural}, that can accommodate heavy SUSY particles in accordance with the Higgs mass. Although the masses of the non-colored particles do not have much of a constraint, but based on the allowed parameter space of the new scenarios, thermal DM is not favored unless the DM particle is very heavy too. For example, within natural SUSY scenarios the Higgsino typically arises as the DM candidate. The annihilation rate in this case is larger than the nominal value for thermal scenario $3\times 10^{-26}$ cm$^3$/sec for sub-TeV DM mass. On the other hand, such large annihilation rates are becoming more and more constrained by the Fermi-LAT data from DM annihilation in the dwarf galaxies and the galactic center~\cite{Fermi}. As a result, one requires non-thermal mechanisms and/or other DM candidates in order to obtain the correct DM abundance.

Another challenge for physics beyond the SM is the explanation of the baryon asymmetry of the universe~\cite{Baryo}. Constraints on the stop mass have also put the electroweak baryogenesis~\cite{EW} in the minimal supersymmetric standard model (MSSM) in a tight corner. There are many alternative scenarios that can explain the matter-antimatter asymmetry like leptogenesis~\cite{Lepto}, Affleck-Dine baryogenesis~\cite{AD}, hidden sector baryogenesis~\cite{Hidden}, etc. One curious observation is that the energy densities in baryons and DM are of the same order of magnitude, the so-called ``baryon-DM coincidence puzzle''. Then a question arises as whether this apparent coincidence may be addressed by an underlying connection between the DM production and baryogenesis scenarios~\cite{Coincidence}.

In this work, we present a minimal extension of the SM to address these questions. We introduce renormalizable baryon number violating interactions in the Lagrangian that can lead to a successful baryogenesis. The minimal field content that is required to achieve this includes iso-singlet color-triplet scalars and one singlet fermion. We show that the fermion becomes stable, hence a DM candidate, when its mass is around ${\cal O}({\rm GeV})$.
The DM relic density and the baryon asymmetry are produced non-thermally from the decay of some heavy particle(s). Non-thermal baryogenesis has the virtue that couplings associated with the new fields do not need to be artificially small. Moreover, the non-thermal mechanism can correlate the DM relic abundance and baryon asymmetry. Since DM mass is ${\cal O}({\rm GeV})$, a correlation between the number densities will automatically translates into a similar relation between the DM and baryon energy densities. This can provide a natural explanation of the baryon-DM coincidence puzzle.

In this model, the DM candidate interacts with up-type quarks via the exchange of colored scalar fields. We see that the resulting spin-independent and spin-dependent DM-nucleon scattering cross sections are well below the bounds from current and upcoming experiments, which makes the prospects for direct detection weak.
In addition, due to its low mass of ${\cal O}({\rm GeV})$, indirect signals from DM annihilation will be negligible. However, the model may be probed at the LHC via the colored scalars if they have ${\cal O}({\rm TeV})$ masses. The decay channel including the DM candidate will give rise to a missing energy signal. The SUSY extension of this model is straightforward. In this case the scalar partner of the DM can also become stable, if $R$-parity is conserved, which allows a scenario with multi-component DM.

This paper is organized as follows. In Section II, we present the model and see how it can lead to a DM candidate that is nearly degenerate with the proton. In Section III, we discuss non-thermal production of DM and baryogenesis in the model, and show how this can naturally address the baryon-DM coincidence problem. In Section IV, we discuss prospects for probing the model via direct detection experiments and at the LHC, and comment on the SUSY extension of the model and the possibility of having a multi-component DM scenario. We conclude the paper in Section V.

\section{The model}

We start with the SM Lagrangian and add renormalizable terms to it that violate baryon number in order to find a successful baryogenesis scenario. Then gauge invariance requires introducing new colored fields. A minimal set up includes two iso-singlet color-triplet scalars $X_\alpha$ ($\alpha=1,~2$) with hypercharge $+4/3$. This allows us to have the baryon number violating interaction terms $X_\alpha d^c d^c$ in the Lagrangian (we have used two-component Weyl fermions).~\footnote{One can also add a term $X u^c d^c$ to the Lagrangian if $X$ has the same gauge charges as $d^c$ (i.e., iso-singlet color triplet with hypercharge $-2/3$). This leads to similar consequences, and hence we do not consider it as a separate case here.}

We note that at least two $X$ are needed to produce a baryon asymmetry from the interference of tree-level and one-loop diagrams in a decay process governed by the $X_\alpha d^c d^c$ interactions. However, although necessary, this is not sufficient. The reason being that the total asymmetry vanishes after summing over all flavors of $d^c$ in the final and intermediate states~\cite{KW}. One therefore requires additional baryon number violating interactions, and the simplest renormalizable term as such is $X^* N u^c$ where $N$ is a SM singlet. This leads to the following Lagrangian
\bea \label{lagran}
{\cal L} & = & {\cal L}_{\rm SM} + {\cal L}_{\rm new} \, \nonumber \\
{\cal L}_{\rm new} & = & (\lambda_{\alpha i} X^*_\alpha N u^c_{i} + \lambda^\prime_{\alpha i j} {X}_\alpha {d^c_i} d^c_j + {1 \over 2} m_N N N + {\rm h.c.})\,  \nonumber \\
& + & m^2_\alpha |X_\alpha|^2 + ({\rm kinetic ~ terms}) \, .
\eea
Here $i,~j$ denote flavor indices (color indices are omitted for simplicity). We note that $\lambda^\prime_{i j}$ is antisymmetric under $i \leftrightarrow j$. We will discuss the generation of baryon asymmetry form this lagrangian in detail in the next section.
The singlet $N$, which plays an important role in baryogensis, can have a gauge charge under a higher ranked symmetry group that includes the SM.

Assuming that $m_N \ll m_\alpha$, one finds an effective four-fermion interaction $N u^c_i d^c_j d^c_k$ after integrating out $X_\alpha$. The existence of this term implies that $N$ decays to three quarks if $m_N \gg {\cal O}({\rm GeV})$. Also, the decay modes $N \rightarrow p + e^{-} + {\bar \nu}_e , ~ N \rightarrow {\bar p} + e^{+} + \nu_e$ are open as long as $m_N > m_p + m_e$, where $m_p$ and $m_e$ are the proton mass and the electron mass respectively.

However, $N$ becomes absolutely stable if $m_N < m_p - m_e$. The important point to note is that stability of $N$ is not related to any new symmetry. All required is that the proton be stable. Then the same symmetry that ensures stability of the proton, combined with the kinematic condition $m_N < m_p - m_e$, will lead to $N$ being a stable particle. This is the first remarkable property of the DM in this model: the field $N$ that is required for baryogenesis also becomes a DM candidate only if its mass is $\ls {\cal O}({\rm GeV})$.

In  addition, we note that the decays $p \rightarrow N + e^{+} + \nu_e ~ ({\bar \nu}_e)$ are kinematically allowed if $m_p > m_N + m_e$. This is unacceptable as it will result in catastrophic proton decay. Therefore a viable scenario arises provided that
%
\be \label{mass}
m_p - m_e \leq m_N \leq m_p + m_e \, .
\ee
%
This is the second remarkable property of the model: the Lagrangian in Eq.~(\ref{lagran}) gives rise to a viable DM candidate if and only if $m_N \approx m_p$.

We therefore see that the model in Eq.~(\ref{lagran}) not only gives rise to baryogenesis, but can also yield a light DM candidate of ${\cal O}({\rm GeV})$ mass. As we will see in the next section, this can lead to a natural explanation of the baryon-DM coincidence problem.

Some comments are in order. First, the smallness of $m_N$ is protected against quantum corrections since $N$ is a fermion. The one-loop corrections arising from the $\lambda X^* N u^c$ coupling result in $\delta m_N \sim (\lambda/4 \pi)^2 {\rm ln} (\Lambda/m_X)$, where $\Lambda$ is a cut off. This implies that even for $\lambda \sim {\cal O}(1)$ we have $\delta m_N \ll m_N$. Therefore the relation $m_N \approx m_p$, once satisfied at the tree-level, will not be destabilized by radiative corrections. In fact, for $\lambda \sim {\cal O}(0.1)$ we have $\delta m_N \ls m_e$.

Second, being a SM singlet Weyl fermion, one may be tempted to identify $N$ with the right-handed (RH) neutrino. This, however, will allow Dirac Yukawa couplings of the neutrino $H N L$ in the Lagrangian. Together with the terms $X^* N u^c$ and $X d^c d^c$, this induces the dimension-7 operator $H L u^c d^c d^c$, which will lead to a rapid proton decay. In order to ensure stability of the proton, one therefore needs to forbid the $H N L$ term. This may be achieved, for example, by introducing a gauged $U(1)_L$ symmetry. Then, similar to the case with a gauged $U(1)_{B-L}$, anomaly cancellation will require the existence of three RH neutrinos. The $U(1)_L$ symmetry does not allow a coupling between the RH neutrinos, which carry lepton number, and the $X^* u^c$ combination. Therefore, the singlet $N$ that participates in the $X^* N u^c$ interaction term will be decoupled from the lepton sector. This guarantees the stability of the proton, which also implies the stability of DM in our model.

Finally, we note that it is possible to make the DM mass much lower than the proton mass by forbidding the $X d^c d^c$ term in the Lagrangian~(\ref{lagran}) by invoking some discrete symmetry. In this case the proton decay constraint does not apply. However, the absence of the $X d^c d^c$ term also implies that there will be no baryogenesis.


\section{Baryogenesis and dark matter production}

In this section, we discuss baryogenesis and production of DM in the model. We see that a single decay process is the non-thermal origin of both and results in comparable abundances of the baryon asymmetry and DM. Considering the DM mass $m_N \approx m_p$, this can naturally address the baryon-DM coincidence puzzle.

\subsection{Generation of the baryon asymmetry}


The model in Eq.~(\ref{lagran}) can give rise to generation of baryon asymmetry through the decay of colored scalars $X_\alpha$.
The interference between tree-level and one-loop self-energy diagrams will result in the following baryon asymmetry per decay of $X_1$ and $X_2$ respectively~\cite{Late}
\bea \label{asymmetry}
\epsilon_{1} & = & {1 \over 8 \pi} ~ {\sum_{i,j,k} {\rm Im} \left(\lambda^*_{1k}\lambda_{2k}\lambda^{\prime *}_{1ij}\lambda^{\prime}_{2ij}\right) \over \sum_{i,j} |\lambda^{\prime}_{1ij}|^2 + \sum_{k} |\lambda_{1k}|^2} ~ {\cal F}_S \left(m^2_{1} \over m^2_{2} \right) \, , \nonumber \\
\epsilon_{2} & = & {1 \over 8 \pi} ~ {\sum_{i,j,k} {\rm Im} \left(\lambda^*_{2k}\lambda_{1k}\lambda^{\prime *}_{2ij}\lambda^{\prime}_{1ij}\right) \over \sum_{i,j} |\lambda^{\prime}_{2ij}|^2 + \sum_{k} |\lambda_{2k}|^2} ~ {\cal F}_S \left(m^2_{2} \over m^2_{1} \right) \, , \nonumber \\
& & \,
\eea
where
\be \label{self}
{\cal F}_S(x) = {x \over x - 1}.
\ee
We note some differences between the asymmetry parameter in this case and that in the leptogenesis scenario~\cite{CRV}. First, there are no one-loop vertex diagrams here. Second, the numerator of the expression for ${\cal F}(x)$ (\ref{self}) contains $x$ instead of $\sqrt{x}$. The reason being that the baryon number violation in the self-energy diagram arises from the couplings $X^* N u^c$ and $X d^c d^c$ in this case, while a Majorana mass term is responsible for the asymmetry in leptogenesis.

The generated baryon asymmetry normalized by the entropy density $s$ is given by
\be
\eta_B \equiv {n_B - n_{\bar B} \over s} = \epsilon_1 {n_{X_1} \over s} + \epsilon_2 {n_{X_2} \over s} .
\ee
%
%
A natural choice of parameters is $|\lambda|,~|\lambda^{\prime}| \sim {\cal O}(1)$ and $CP$ violating phases of ${\cal O}(1)$, which for $m_1 \sim m_2$ results in $\epsilon_{1,2} \sim {\cal O}(0.1)$. However, for $m_{1,2} \sim {\cal O}({\rm TeV})$, the third Sakharov condition~\cite{Sakharov} for generating a baryon asymmetry (i.e., out-of-equilibrium decay of $X_1,~X_2$) requires that $|\lambda|,~|\lambda^{\prime}| \ll 1$ if thermal initial condition for $X_1,~X_2$ is assumed. Moreover, in this picture, the comoving number density of $X_1,~X_2$ exponentially decreases due to their annihilation into gluons as the temperature drops below $m_{1,2}$. This implies that $n_{X_1},~n_{X_2}$ are too suppressed at the time of decay to yield the desired baryon asymmetry $\eta_B \sim {\cal O}(10^{-10})$.

Therefore, a non-thermal scenario is needed for successful realization of baryogenesis in this model. In a possible scenario $X_1$ and $X_2$ are produced from the late decay of a scalar field $S$ with mass $m_S$ that reheats the universe to a low temperature $T_r$~\cite{Late}.
%
Then the overall baryon asymmetry will be
\bea \label{etaB}
&& \eta_B ~ \sim ~ Y_S ~\cdot {1 \over 8 \pi} \cdot {1 \over {m^2_1 - m^2_2}} ~ \sum_{i,j,k} {\rm Im} \left(\lambda^*_{1k}\lambda_{2k}\lambda^{\prime *}_{1ij}\lambda^{\prime}_{2ij}\right) \times \, \nonumber \\
\, \nonumber \\
&& \left[{m^2_1 ~ {\rm Br}_1 \over \sum_{i,j}|\lambda^{\prime}_{1ij}|^2 + \sum_{k} |\lambda_{1k}|^2} + {m^2_2 ~ {\rm Br}_2 \over \sum_{i,j} |\lambda^{\prime}_{2ij}|^2 + \sum_{k} |\lambda_{2k}|^2} \right] \, . \nonumber \\
\,
\eea
Here $Y_S \equiv 3 T_{\rm r}/4 m_S$ is the dilution factor due to entropy release by the late decaying scalar field $S$, and ${\rm Br}_{1,2}$ denote the branching ratios for producing $X_1$ and $X_2$ from $S$ decay respectively.

\subsection{Non-thermal production of dark matter}

The DM candidate $N$ reaches equilibrium with the primordial plasma at sufficiently high temperatures through its interactions with $u^c$ given in Eq.~(\ref{lagran}). At temperatures $T \ll m_{1,2}$, the cross section for $N$ scattering off $u^c$, or its pair creation from quark-antiquark annihilations, is given by $\sigma \sim \lambda^4 T^2/m^4_{1,2}$. The corresponding interaction rate exceeds the Hubble expansion rate provided that $T \gs (\lambda^{-4} m^{4}_{1,2}/M_{\rm P})^{1/3}$. For $\lambda \sim {\cal O}(1)$ and $m_{1,2} \sim {\cal O}({\rm TeV})$, this implies rapid equilibration of $N$ at temperatures as low as $T \sim m_N \approx {\cal O}({\rm GeV})$.
%
The comoving number density of $N$ decreases due to pair annihilation as the temperature drops below $m_N$ until thermal freeze-out of $N$ annihilation. However, for $m_N \approx m_p$ and $m_{1,2} \sim {\cal O}({\rm TeV})$, thermal freeze-out leads to overabundance of $N$ according to the Lee-Weinberg bound~\cite{LW}.

Therefore obtaining the correct relic abundance for $N$ requires a non-thermal scenario of DM production. Late decay of a scalar $S$ that
reheats the universe to a low temperatures $T_{\rm r} \ll {\cal O}({\rm GeV})$ can be the origin of non-thermal DM production. Interestingly, as we argued above, such a scenario is also required for successful baryogenesis in this model.
In fact, the same decay processes that generate baryon asymmetry also produce the DM candidate $N$. The relic density of DM particles thus produced is given by
\bea \label{dmdens}
&& {n_N \over s} ~ \sim ~ Y_S \times \, \nonumber \\
&& \left[{{\rm Br}_1 ~ \sum_{k}|\lambda_{1k}|^2 \over \sum_{i,j}|\lambda^{\prime}_{1ij}|^2 + \sum_{k}|\lambda_{1k}|^2} + {{\rm Br}_2 ~ \sum_{k}|\lambda_{2k}|^2 \over \sum_{i,j}|\lambda^{\prime}_{2ij}|^2 + \sum_{k}|\lambda_{2k}|^2} \right] \, . \nonumber \\
&&
\eea

$N$ quanta, although produced when the temperature of the plasma is $T_{\rm r} \ll m_N$, are ultra relativistic since their initial energy is set by the mass of decaying particle $m_S$. To determine whether $N$ is cold or warm DM, one needs to see whether $N$ can reach kinetic equilibrium with the plasma quickly enough. The interactions of $N$ with $u^c$, via exchange of $X_{1,2}$, also couple $N$ to $\pi$'s. At temperatures $m_\pi/3 \ls T_{\rm r} \ll {\cal O}({\rm GeV})$, $\pi$'s are abundant and kinetic equilibrium of $N$ with the plasma is achieved through its interactions with $\pi$'s. At lower temperatures, $T_{\rm r} < m_\pi/3$, the number density of $\pi$'s is exponentially suppressed. In this case $N$ interaction with the plasma occurs mainly through $\pi$ exchange with the photons.

We have checked that kinetic decoupling temperature of $N$ is $\sim {\cal O}({\rm MeV})$, which is similar to the neutralino DM~\cite{Kinetic}, if $m_{1,2},~m_S \sim {\cal O}({\rm TeV})$ and $\lambda \sim {\cal O}(1)$. Achieving kinetic equilibrium with the plasma very rapidly makes $N$ a cold DM candidate. For much larger values of $m_{1,2},~m_S$ and/or much smaller values of $\lambda$, kinetic decoupling temperature will be much higher. If it exceeds $T_{\rm r}$, kinetic equilibrium of $N$ will not be achieved, in which case $N$ can be warm DM.

\subsection{Explanation of the baryon-DM coincidence problem}

The interesting point to note from Eqs.~(\ref{etaB},\ref{dmdens}) is that the DM and baryon number densities have very similar functional dependence on the model parameters. As a result, they can come within the same order of magnitude without making very special assumptions. For the natural choice of parameters $|\lambda| \sim |\lambda^{\prime}| \sim {\cal O}(1)$, and $CP$-violating phases of ${\cal O}(1)$, one indeed finds that $n_N/s$ is larger than $\eta_B$ by a factor of a few. Since $m_N \approx m_p$, this is directly translated into similar relation between the DM and baryon energy densities, thus providing a natural explanation for the DM-baryon coincidence puzzle.

The model in Eq.~(\ref{lagran}) therefore leads to a successful realization of the ``Cladogenesis'' scenario~\cite{Clado}. In this scenario the DM and baryon densities are mainly controlled by the dilution factor and branching ratios of the decaying field $S$. Assuming ${\cal O}(1)$ values for $\lambda$ and $\lambda^{\prime}$, one finds $\eta_B$ and $n_N/s$ in the ballpark of measured values
due to the smallness of $Y_S$ and ${\rm Br}_{1,2}$.
The exact numbers can then be obtained by minor adjustments of $\lambda$ and $\lambda^{\prime}$ about their natural values.

One can construct explicit models of a late-decaying scalar field $S$ that has a small branching ratio to certain fields as discussed in~\cite{Higgsino,Visible}. In the case at hand, for example, one can assign lepton number $-2$ to $S$ within a gauged $U(1)_L$ model. This allows $S$ coupling to the RH neutrino $\nu_R$ via a lepton number conserving term $h S \nu_R \nu_R$. As usual, $\nu_R$ participates in the neutrino Dirac Yukawa coupling terms $h_\nu H \nu_R L$. The $U(1)_L$ symmetry allows renormalizable couplings of the form $|S|^2 |X_{1,2}|^2$ between $S$ and $X_{1,2}$, but not terms like $S N N$. Assuming that $m_{\nu_R} < m_S < 2 m_{\nu_R}$, the main decay mode of $S$ is $S \rightarrow H \nu_R L$. Here $m_{\nu_R}$ is the Majorana mass of the RH neutrino, which is generated after spontaneous breaking of $U(1)_L$ by some Higgs field. The $S$ decay rate is suppressed $\propto (h h_\nu)^2$, in addition to the three-body phase space suppression. This can explain a small decay rate leading to a low reheat temperature $T_{\rm r}$, and hence a small value of $Y_S$. Production of $X_{1,2}$ occurs via 5-body final state decays $S \rightarrow X_{1,2} X^*_{1,2} H \nu_R L$, which result in very small values of ${\rm Br}_{1,2}$.



\section{Direct detection and collider signals}

The DM candidate $N$ interacts with nucleons through its coupling to the $u^c$ mediated by $X$, see Eq.~(\ref{lagran}). However, $N$ only couples to a particular chirality of up-type quarks. As a result, there are no effective interactions of the form $(\bar \psi_N) \psi_N ({\bar \psi_q} \psi_q)/m^2_X$ between the DM and quarks ($\psi_N$ and $\psi_q$ are four-component fermions representing $N$ and quark fields respectively). The spin-independent interactions with nucleons arise from twist-2 quark operators and one-loop diagrams that couple $N$ to gluons~\cite{DN}. The corresponding elastic scattering cross section will be suppressed $\propto m_X^{-8}$, which gives rise to $\sigma_{\rm SI} \ls 10^{-16}-10^{-15}$ pb for $m_X \sim {\cal O}({\rm TeV})$. This is considerably below the reach of upcoming experiments.

The spin-dependent cross-section is only suppressed $\propto m^{-4}_X$ as one has effective interactions of the form $({\bar \psi_N} \gamma_5 \gamma^\mu \psi_N) ({\bar \psi_q} \gamma_5 \gamma_\mu \psi_q)/m^2_X$. This results in $\sigma_{\rm SD} \ls 10^{-6}-10^{-5}$ pb, for $m_X \sim {\cal O}({\rm TeV})$, which is much below the bounds from current experiments~\cite{KIMS}, as well as the upcoming detection experiments. It is also significantly below the current LHC bounds on $\sigma_{\rm SD}$~\cite{Fox}, but may be within the LHC future reach~\cite{Tait}.

Indirect detection signals from DM annihilation will also be negligible in this model. The reason being that because of the low DM mass, any signal will be overwhelmed by the astrophysical background.

DM interactions with matter have a novel signature in this model that may be seen. The effective interaction $N u^c d^c d^c$ leads to baryon destroying inelastic scattering of $N$ off nucleons, similar to the model in~\cite{Hylo}, which may have an appreciable rate for nucleon decay experiments.

Moreover, this model can be observed at the LHC. The colored scalars $X_{1,2}$ can be pair produced and subsequently decay to a $d^c d^c$ pair or to $N u^c$, where $N$ constitutes missing energy. The final state of $X$ pair production can be 4 jets, 3 jets plus missing energy, 2 jets plus missing energy, etc. It is interesting to note that the final state with 4 jets does not involve any missing energy.

Extending this model to a SUSY version is straightforward and brings further interesting prospects. A minimal extension is done in the context of MSSM by introducing the SUSY partners of $X$ scalars and $N$ fermion, denoted by ${\tilde X}$ and ${\tilde N}$ respectively, plus two iso-singlet color-triplet superfields ${\bar X}_{1,2}$ with hypercharge $Y = -4/3$ required by anomaly cancellation considerations. The Lagrangian ${\cal L}_{\rm new}$ in Eq.~(\ref{lagran}) will be elevated to a superpotential $W_{\rm new}$ to include all these fields and their interactions.

If $R$-parity is conserved, then the LSP, denoted by $\chi$, will be stable and can be a DM candidate as well. This leads to the possibility of multi-component DM consisting of $N$ and $\chi$. The LSP component of DM may be detected via direct searches even if it is subdominant. For example, $\chi$ can be the Higgsino, or the superpartner of $N$~\cite{Rabi2}, both of which have elastic scattering cross sections within the reach of upcoming experiments. The multi-component DM scenario may also be detected via indirect searches. For example, a subdominant Higgsino component can lead to a detectable gamma-ray signal that is compatible with the Fermi-LAT bounds~\cite{Fermi}. In this case, concordance between the collider, direct, and indirect signals would imply the necessity of a dominant non-LSP component of DM. However, complementary signals from direct and indirect searches are more constrained due to the subdominance of the LSP component~\cite{BB}.

In a SUSY set up one has to ensure that $S$ decay does not lead to overproduction of the LSP component of DM. This can be achieved
, for example, if $S$ is an $R$-parity even scalar that belongs to the visible sector~\cite{Visible}.
We note that in $R$-parity violating models $N$ is the only DM candidate. In this case production of $R$-parity odd particles does not need to be suppressed since the LSP is not stable.

At the LHC, the $R$-parity odd colored fields $\tilde X$ are pair produced and then decay to ${\tilde d}^c d^c$, ${\tilde u}^c N$, $ u^c {\tilde N}$ final states. All of these final states involve jets plus missing energy. Also, if $\tilde X$ is lighter than squarks, then the squark decay mode may contain the information of $\tilde X$. For example, ${\tilde u}^c \rightarrow X + \tilde N,~\tilde X + N$, with $\tilde N,~\tilde X$ subsequently decaying into the LSP. The $R$-parity even colored fields $X$ decay to $d^c d^c$, $ N u^c$, ${\tilde d}^c {\tilde d}^c$, ${\tilde N} {\tilde u}^c$ final states, with missing energy arising from $N$ or from the LSP. An important difference is that the SUSY version of the model contains both 4 jets and 4 jets plus missing energy final states.

Finally, let us briefly mention some of the other phenomenological implications of this model. The $X d_i^c d_j^c$ coupling can lead to $K^0_s$-$\bar K^0_s$ or $B^0_s$-$\bar B^0_s$ mixing. However, color conservation does not allow any tree-level contribution to the mixing term, and  one-loop contributions that generate the relevant operators satisfy the experimental constraints easily~\cite{Late}.
This model also gives rise to neutron-antineutron oscillations through the dimension-9 operator $G \propto (u^c d^c s^c)^2/(m^4_X m_N)$.
The oscillation time scale is given by $t \sim 1/(2.5 10^{-5}\, G)$ s, where $2.5 \times 10^{-5}$ is the value for the hadronic form factor~\cite{shrock}. The current bound $t < 0.86 \times 10^8$ sec~\cite{expt} requires that $G < 3 \times 10^{-28}$ ${\rm GeV}^{-5}$. Since all possible flavor combinations appear in the expression for baryon asymmetry, see Eq.~(\ref{etaB}), it is easy to satisfy the oscillation bound while obtaining the correct baryon asymmetry~\cite{Late}.

\section{Conclusion}

In conclusion, we considered a simple extension of the SM that gives rise to baryogenesis and has a DM candidate of ${\cal O}({\rm GeV})$ mass. Two colored scalars with ${\cal O}({\rm TeV})$ mass and a singlet fermion are required in a minimal set up to generate the baryon asymmetry of the universe via renormalizable baryon number violating interactions. The singlet fermion becomes stable and can play the role of a DM candidate, while avoiding rapid proton decay, when it is nearly degenerate in mass with the proton. None of these explanations requires the existence of SUSY.

DM and baryon asymmetry are produced non-thermally form out-of-equilibrium decay of new colored fields. Non-thermal baryogenesis is motivated when couplings associated with the new fields are not chosen to be artificially small. Obtaining the correct DM relic abundance for an ${\cal O}({\rm GeV})$ particle that interacts with matter via exchange of ${\cal O}({\rm TeV})$ scale fields also requires a non-thermal scenario.
This mechanism correlates the DM abundance and baryon asymmetry for natural choice of parameters.
Since DM mass is ${\cal O}({\rm GeV})$, a correlation between the number densities is directly translated into a similar relation between the energy densities. Therefore this model provides a natural explanation of the baryon-DM coincidence puzzle.

The model predicts
DM-nucleon scattering cross sections $\sigma_{\rm SI} \ls 10^{-16}-10^{-15}$ pb and $\sigma_{\rm SD} \ls 10^{-6}-10^{-5}$ pb are far below the bounds from current and upcoming direct detection experiments. The predicted $\sigma_{\rm SD}$ may however be within the LHC future reach. The indirect signals from DM annihilation are also negligible due to its low mass. However, the model may be probed at the LHC via the colored scalars. Pair production and subsequent decay of these particles results in final states with $N$ that constitutes missing energy. Final states with 4 jets, 3 jet plus missing energy, and 2 jet plus missing energy can arise but, interestingly, there will be no final states with 4 jets plus missing energy.

Extending this model to a SUSY version is straightforward. In $R$-parity conserving models the LSP, which may be the scalar partner of $N$, is also stable and can be a DM candidate too. This allows a scenario with multi-component DM, which has a better prospect for direct and indirect detection experiments. In the SUSY extension, final states with 4 jets with missing energy can also arise.

\section{Acknowledgement}

The work of B.D. is supported in part by the DOE grant DE-FG02-95ER40917.


\end{document}